\documentstyle[twoside,fleqn,npb,epsfig]{article}

\newcommand{\AmS}{{\protect\the\textfont2
  A\kern-.1667em\lower.5ex\hbox{M}\kern-.125emS}}

\title{Jet Physics at Two-Loop Accuracy
\thanks{Presented at {\it RADCOR2002 / Loops and Legs 
in Quantum Field Theory},
September 2002, Kloster Banz, Germany}
}

\author{T.\ Gehrmann 
\address{Institut 
f\"ur Theoretische Physik,
RWTH Aachen, D-52056 Aachen, Germany}}

\begin{document}
\unitlength 1cm

\begin{abstract}
Current phenomenological studies of jet observables at colliders are 
clearly limited by the theoretical uncertainties inherent in the 
next-to-leading order QCD description.  
We discuss the recent progress made towards the calculation of 
QCD corrections to jet observables at the 
next-to-next-to-leading order in QCD and highlight future perspectives and
yet open issues.

\end{abstract}

\maketitle

\section{Introduction}
Jet production observables are among the most sensitive probes of QCD at
high energy colliders, where they are used for example to determine the
strong coupling constant. At present, the interpretation of jet production
data within perturbative QCD is restricted to next-to-leading 
order (NLO)
calculations, with theoretical uncertainties considerably larger
than current experimental errors. Going beyond NLO calculations offers moreover
a more accurate matching of theoretical and experimental jet definitions and 
a more detailed modelling of the hadronic final state~\cite{nigel}.
The extension of jet calculations to
NNLO requires three ingredients: the two-loop corrections to
multi-leg amplitudes, the single unresolved limits of one-loop
amplitudes and the double unresolved limits of tree
amplitudes. Finally, all these contributions have to be combined
together into a program for the numerical computation of jet
observables from NNLO parton level cross sections. 
In this talk, I review recent
progress made on these subjects as well as the currently open issues.

\section{Virtual Two-Loop Corrections}
Within dimensional
regularization,  the large number
of different integrals appearing in  
multi-loop calculations can be reduced
to a small number of  so-called {\em master integrals} by using
integration-by-parts (IBP)
identities~\cite{hv,chet}. These identities
exploit the fact that the integral over the total derivative of 
any  of the loop
momenta vanishes in dimensional regularization.

For integrals involving more than two external legs,  another
class of identities exists due to  Lorentz invariance.
 These
Lorentz invariance identities~\cite{gr} rely on the fact that  an
infinitesimal Lorentz transformation commutes with the loop integrations, thus
relating different integrals. The common origin of IBP and LI 
identities is the Poincare invariance of loop integrals within
dimensional regularization, as was pointed out by J.J.\ van der Bij at
this conference.
Using integration-by-parts  and Lorentz
invariance identities, all  two-loop Feynman amplitudes for $2\to 2$
 scattering or $1\to 3$ decay  processes  
can be expressed as linear combinations of a
small number of  master integrals,
which have  to be computed by some different method.  Explicit
reduction formulae for on-shell two-loop four-point  integrals were derived
in~\cite{onshell3}. 
Computer algorithms for the  automatic reduction of all
two-loop four-point integrals  were 
described in~\cite{gr,laporta}. 

The master integrals relevant to $2\to 2$ scattering or $1\to 3$ decay
processes are massless, scalar two-loop  four-point functions with all legs
on-shell or a single leg off-shell.  Several techniques for the 
computation of
those functions  have been proposed in the literature, 
such as the application
of a Mellin-Barnes transformation to all 
propagators, which was used 
successfully to compute the on-shell planar double box 
integral~\cite{onshell1,onshell6} , the on-shell non-planar double box
integral~\cite{onshell2} and two double box integrals with one leg 
off-shell~\cite{smirnov1}. Most recently, the same method was 
used to  derive the on-shell planar double box integral with one 
internal mass scale~\cite{smirnov2} as well as the high energy limit of the 
on-shell planar triple box integral~\cite{smirnov3}. 
\begin{figure*}[tbh]
\epsfig{file=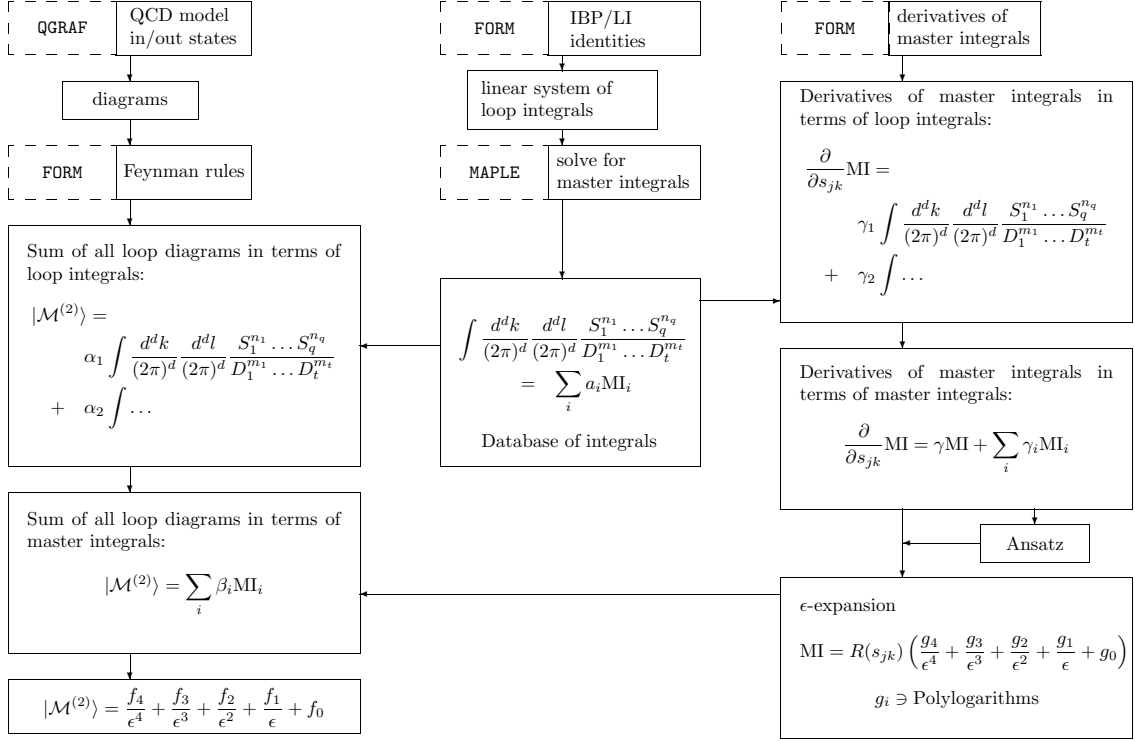,angle=-90,width=15cm}
\caption{Computer Algebra for analytic evaluation of two-loop matrix elements.}
\label{fig:algebra}
\end{figure*}

A method for the analytic computation of master integrals avoiding the 
explicit
integration over the loop momenta is to derive differential equations in
internal propagator masses or in external momenta for the master
integral,  
and
to solve these with appropriate boundary conditions.   The
computation of master integrals from differential equations proceeds
as  follows~\cite{gr}. Carrying out the derivative with respect to an 
external invariant  on
the master integral of a given topology, one obtains a linear
combination 
of a
number of  more complicated integrals, which can however be reduced to the 
master integral itself plus simpler integrals by applying the reduction 
methods discussed above. As a result, one obtains  an inhomogeneous linear
first order differential equation in each invariant for the master integral.
The inhomogeneous term in these differential equations contains only 
topologies simpler than the topology under consideration, which are 
considered
to be known  if working in a bottom-up approach.  
The master integral is then
obtained by matching the general  solution of its differential
equation 
to an
appropriate boundary condition. 

Using the differential equation technique, one of the on-shell 
planar double box integrals~\cite{onshell5} as well as the full set of
planar and non-planar off-shell double box 
integrals~\cite{mi}
 were derived.  The computer algebra structures applied in the 
computation of the master integrals from the differential equations 
are displayed in the right hand column of Figure~\ref{fig:algebra}.
The differential equation approach has recently been extended to 
phase space integrals~\cite{babis}.

A strong check on all these computations of master integrals is  given
by the
completely numerical calculations of~\cite{num}, 
which are  based on an iterated
sector decomposition to isolate the infrared pole  structure. 
The methods
of~\cite{num} were applied to 
confirm {\em all}  of the above-mentioned
calculations. 

A third approach, which avoids the reduction to master integrals, has
been presented in~\cite{muw1}. In this approach, all integrals 
appearing in the two-loop amplitudes are related to higher
transcendental functions, which can be expanded in terms of nested 
harmonic sums.

The two-loop four-point functions with all legs on-shell can be 
expressed  in
terms of Nielsen's 
polylogarithms~\cite{nielsen,bit}.  In
contrast, the closed analytic  expressions for  two-loop four-point functions
with one leg off-shell contain  two new classes of functions: harmonic 
polylogarithms~\cite{hpl} and two-dimensional
harmonic  polylogarithms (2dHPL's)~\cite{mi}.  Accurate 
numerical
implementations for these 
functions~\cite{1dhpl} are
available. 

\subsection{$2 \to 2$ Processes with all legs on-shell}

With the explicit solutions of the integration-by-parts and  Lorentz-invariance
identities for on-shell two-loop four-point
functions and the
corresponding master  
integrals, all
necessary ingredients  for the 
computation of two-loop corrections to  $2 \to 2$
processes with all legs on-shell are now available. The generic
structure of such a calculation is outlined in the left hand column of
Figure~\ref{fig:algebra}.
In fact, only half a  year
elapsed between the completion of the full set of master 
integrals and the calculation of the 
two-loop QED corrections to Bhabha-scattering~\cite{m1}.  
Subsequently,
results were obtained for the two-loop  QCD corrections to all parton-parton
scattering processes~\cite{m2}. For gluon-gluon scattering, the 
two-loop helicity amplitudes have
also  been derived~\cite{m3}.
Moreover, two-loop corrections were
derived  to processes involving two partons and two real
photons~\cite{m4}. Since the gluon fusion into photons has a vanishing 
tree level amplitude, these results form part of the NLO corrections to 
photon pair production~\cite{bern}, yielding a sizable correction.

Finally, light-by-light scattering in
two-loop QED and QCD was  considered in~\cite{m5}.

The results for the two-loop QED matrix element for  Bhabha
scattering~\cite{m1} were used in~\cite{bhabha} 
to extract the  single
logarithmic contributions to the Bhabha scattering cross section.

\subsection{$2 \to 2$ Processes with one off-shell leg}

Using the two-loop master integrals with one off-shell leg~\cite{mi}, 
the two-loop QCD matrix element for  $e^+e^- \to
3$~jets~\cite{3jme} and the  corresponding helicity 
amplitudes~\cite{3jtensor} were computed following the reduction procedure 
depicted in Figure~\ref{fig:algebra}. The infrared pole structure 
of these results agrees with the prediction~\cite{catani} obtained 
from an infrared factorization formula. 

An independent confirmation of part of these results was performed 
in~\cite{muw2}, where two of the seven colour factors (corresponding to 
the terms proportional to $n_f$) of the two-loop helicity amplitudes 
for  $e^+e^- \to 3$~jets were derived using the nested sum method 
of~\cite{muw1}.

Processes related to  $e^+e^- \to 3$~jets  by crossing symmetry  are
$(2+1)$-jet production in  deep inelastic $ep$ scattering and
vector-boson-plus-jet production at hadron  colliders.  
The analytic continuation of the 
$e^+e^- \to 3$~jets two-loop helicity amplitudes  to the kinematic regions 
relevant
for these scattering processes has been derived in~\cite{ancont}.

\section{Real Corrections}
Besides the two-loop virtual corrections, a NNLO calculation of jet
observables has to include the contributions from single unresolved 
(soft or collinear) real 
radiation from one-loop processes as well as from double unresolved
real radiation at tree level. 
Only after summing all these contributions (and including  terms from the
renormalization of parton distributions  for processes with partons in the
initial state), do the divergent terms cancel  among one another.   The
factorization properties of both the one-loop, one-unresolved-parton
contribution~\cite{onel} and the tree-level, 
two-unresolved-parton contributions~\cite{twot}
have been 
studied,  but a systematic
procedure for isolating the infrared singularities has so far been
established only for the  one-loop, one-unresolved-parton processes. 
Although this is still an open and highly non-trivial issue, significant
progress is anticipated in the near future. 
\begin{figure*}[htb]
\epsfig{file=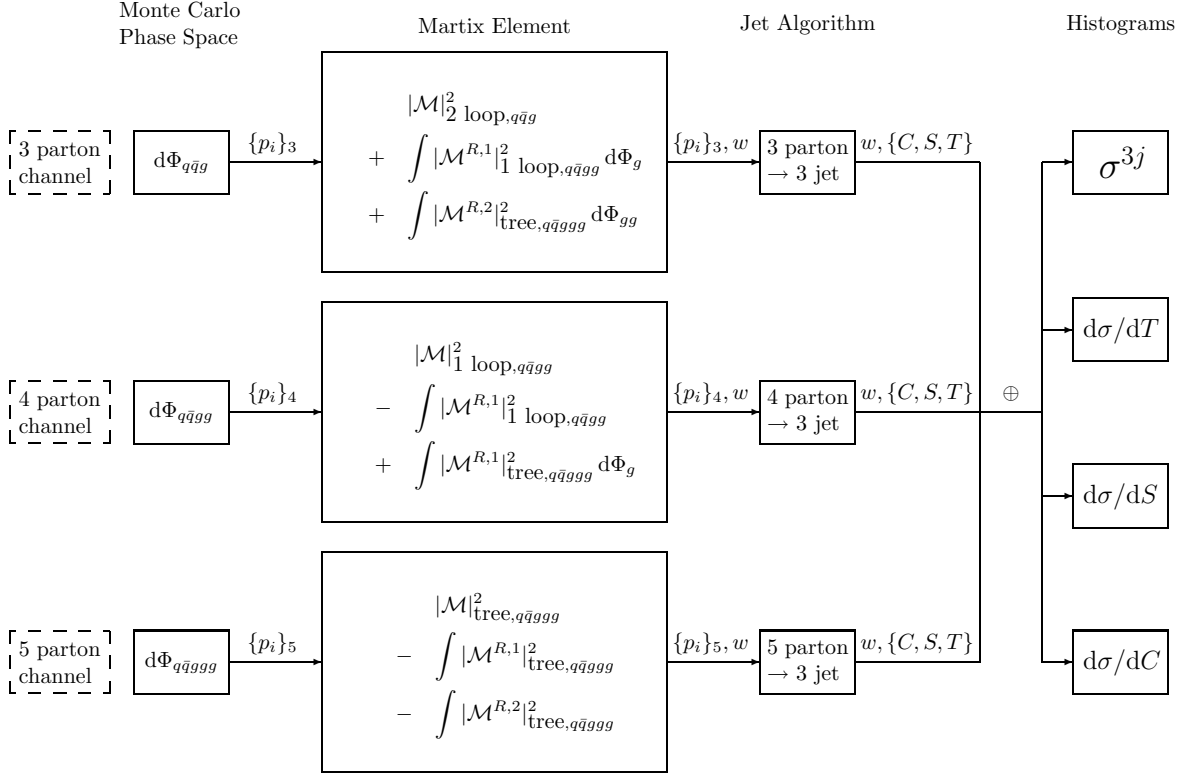,angle=-90,width=16cm}
\caption{Structure of NNLO parton level Monte Carlo programme.}
\label{fig:mc}
\end{figure*}

\section{Numerical Implementation}
Prior to their implementation into a numerical program, it is needed to 
analytically extract the infrared pole terms from the 
one-loop, one-unresolved-parton and two-unresolved-parton contributions. At 
NLO, two types of methods have been used very successfully in the past. The 
{\it phase space slicing} method~\cite{kramer} divides up the final state
phase space into resolved and unresolved regions; the infrared subtraction 
terms are then integrated only over the unresolved regions. This method 
avoids overcounting of singular contributions, the required integrals over
restricted phase space regions can however be very involved beyond NLO. 
The {\it subtraction} method~\cite{ert} integrates the subtraction terms 
over the full phase space. The construction
of the subtraction term requires in this case
great care to avoid overcounting problems. 
In general, algorithms based on subtraction are more efficient numerically.

The remaining finite terms must then be combined  into a numerical program
implementing the  experimental definition of jet observables and 
event-shape variables.  The sketch of such a programme 
to compute $e^+e^- \to 3j$ at NNLO is given in 
Figure~\ref{fig:mc}. 

Programs to compute processes with initial state hadrons 
involve the additional complication of initial state singularities, which 
have to be absorbed into the NNLO parton distributions. Lacking the full 
expressions for the splitting functions at this order, these are not yet 
available at present, work on them is however well advanced~\cite{josnew}.

A first calculation involving the features of NNLO jet calculations 
was presented for
the case of  photon-plus-one-jet final  states in electron--positron
annihilation in~\cite{ggamma}, 
thus demonstrating the feasibility of this type
of calculations. A prerequisite for such a numerical program 
computing $n$ jet final states 
is a stable and
efficient next-to-leading order
programme for the processes yielding $n+1$ jet final states. For the
processes of highest phenomenological interest, these are already
available: $e^+e^- \to 4j$~\cite{4jnum}, $ep\to (3+1)j$~\cite{nagy1},
$pp\to 3j$~\cite{nagy2}, $pp\to V+2j$~\cite{mcfm}.

\section{Conclusions and Outlook}
Considerable progress has been made in the last two years (in fact 
since the last ``Loops and Legs''- and ``RADCOR''-conferences) towards 
the computation of jet observables at NNLO in QCD. In particular, 
new methods have been developed for the calculation of 
two-loop virtual corrections to four-point scattering amplitudes. As a
result, the two-loop virtual corrections relevant to all phenomenologically 
important processes in QCD and QED are now known. 

These form however only part of the full calculation required at NNLO accuracy,
which also has to take into account contributions from one-loop 
single-unresolved radiation and tree-level double unresolved radiation 
processes. While appropriate subtraction terms for these processes have
been known for quite some time, their analytic integration (required 
for the cancellation of infrared poles in physical jet observables) is still 
an unsolved problem. Once this obstacle has been overcome, the remaining 
finite parts can be implemented into a numerical programme to 
compute NNLO jet observables. 

\section*{Acknowledgement}
I wish to thank 
Ettore Remiddi, Nigel Glover, Lee Garland and 
Thanos Koukoutsakis  for a pleasant and fruitful collaboration on the topics 
discussed in this talk.


\begin{thebibliography}{99}


\bibitem{nigel}
E.W.N.~Glover, these proceedings.

\bibitem{hv}
G.\ 't Hooft and M.\ Veltman, Nucl.\ Phys.\ {\bf B44} (1972) 189.

\bibitem{chet}
F.V.\ Tkachov, Phys.\ Lett.\ {\bf 100B} (1981) 65;
K.G.\ Chetyrkin and F.V.\ Tkachov, Nucl.\ Phys.\ {\bf B192} (1981) 159.

\bibitem{gr} 
T.\ Gehrmann and E.\ Remiddi, Nucl.\ Phys.\
{\bf B580} (2000) 485.


\bibitem{onshell3}
V.A.\ Smirnov and O.L.\ Veretin,  Nucl.\ Phys.\ {\bf B566}
(2000) 469;
C.~Anastasiou, E.W.N.~Glover and C.~Oleari,
Nucl.\ Phys.\  {\bf B575} (2000) 416; {\bf B585} (2000) 763(E);
C.\ Anastasiou, T.\ Gehrmann, C.\ Oleari, E.\ Remiddi and 
J.B.\ Tausk, Nucl.\ Phys.\
{\bf B580} (2000) 577.

\bibitem{laporta}
S.~Laporta,
Int.\ J.\ Mod.\ Phys.\ {\bf A15} (2000) 5087.


\bibitem{onshell1}
V.A.\ Smirnov, Phys.\ Lett.\ {\bf B460} (1999) 397.
\bibitem{onshell6}
C.\ Anastasiou, J.B.\ Tausk and M.E.\ Tejeda-Yeomans, 
Nucl.\ Phys.\ {\bf B} (Proc.\ Suppl.) {\bf 89} (2000) 262.
\bibitem{onshell2}
J.B.\ Tausk, Phys.\ Lett.\ {\bf B469} (1999) 225.

\bibitem{smirnov1}
V.A.~Smirnov, Phys.\ Lett.\ {\bf B491} (2000) 130; 
{\bf B500} (2001) 330.

\bibitem{smirnov2}
V.A.~Smirnov, Phys.\ Lett.\ {\bf B524} (2002) 129.

\bibitem{smirnov3}
V.A.~Smirnov, hep-ph/0209193 and these proceedings,
hep-ph/0209295.

\bibitem{onshell5}
T.~Gehrmann and E.~Remiddi, Nucl.\ Phys.\ {\bf B} (Proc.\ Suppl.)
{\bf 89} (2000) 251.

\bibitem{mi}
T.~Gehrmann and E.~Remiddi, Nucl.\ Phys.\ {\bf B601} (2001) 248; 
{\bf B601} (2001) 287.

\bibitem{babis}
C.~Anastasiou and K.~Melnikov,
hep-ph/0207004 and these proceedings.



\bibitem{num}
T.\ Binoth and G.\ Heinrich, Nucl.\ Phys.\ {\bf B585} (2000) 741 and
these proceedings.

\bibitem{muw1}
S.~Moch, P.~Uwer and S.~Weinzierl,
J.\ Math.\ Phys.\  {\bf 43} (2002) 3363;
S.~Weinzierl, Comput.\ Phys.\ Commun.\  {\bf 145} (2002) 357.


\bibitem{nielsen} N. Nielsen, {\it Der Eulersche Dilogarithmus und seine 
Verallgemeinerungen}, Nova Acta Leopoldina (Halle) {\bf 90} (1909) 123;
L.~Lewin, {\it Polylogarithms and
Associated Functions} (North Holland, Amsterdam 1981);
K.S.\ K\"olbig, SIAM J.\ Math.\ Anal.\ {\bf 17} (1986) 1232. 

\bibitem{bit} 
K.S.\ K\"olbig, J.A.\ Mignaco and E.\ Remiddi, BIT {\bf 10} (1970) 38. 

\bibitem{hpl} 
E.\ Remiddi and J.A.M.\ Vermaseren, Int.\ J.\ Mod.\ Phys.\ {\bf A15} 
(2000) 725. 

\bibitem{1dhpl}
T.~Gehrmann and E.~Remiddi,
Comput.\ Phys.\ Commun.\ {\bf 141} (2001) 296; {\bf 144} (2002) 200.



\bibitem{m1}
Z.~Bern, L.~Dixon and A.~Ghinculov, Phys.\ Rev.\ {\bf D63} (2001) 053007.

\bibitem{m2}
C.\ Anastasiou, E.W.N.~Glover, C.\ Oleari and M.E.\ Tejeda-Yeomans,
Nucl.\ Phys.\ {\bf B601} (2001) 318; 
{\bf B601} (2001) 347; {\bf B605} (2001) 486; 
E.W.N.~Glover, C.~Oleari and M.E.~Tejeda-Yeomans, 
Nucl.\ Phys.\ {\bf B605} (2001) 467.


\bibitem{m3}
Z.~Bern, A.~De Freitas and L.~Dixon,
JHEP {\bf 0203} (2002) 018.

\bibitem{m4}
Z.~Bern, A.~De Freitas and L.~J.~Dixon,
JHEP {\bf 0109} (2001) 037 and these proceedings;
C.~Anastasiou, E.W.N.~Glover and M.E.~Tejeda-Yeomans,
Nucl.\ Phys.\  {\bf B629} (2002) 255.




\bibitem{bern}
Z.~Bern, L.~Dixon and C.~Schmidt, hep-ph/0206194 and these proceedings.


\bibitem{m5}
Z.~Bern, A.~De Freitas, L.J.~Dixon, A.~Ghinculov and H.L.~Wong,
JHEP {\bf 0111} (2001) 031.



\bibitem{bhabha}
E.W.N.~Glover, J.B.~Tausk and J.J.~van der Bij,
Phys.\ Lett.\  {\bf B516} (2001) 33.

\bibitem{3jme}
L.W.~Garland, T.~Gehrmann, E.W.N.~Glover, A.~Koukoutsakis and E.~Remiddi,
Nucl.\ Phys.\  {\bf B627} (2002) 107.

\bibitem{3jtensor}
L.W.~Garland, T.~Gehrmann, E.W.N.~Glover, A.~Koukoutsakis and E.~Remiddi,
Nucl.\ Phys.\  {\bf B642} (2002) 227.

\bibitem{catani}
S.\ Catani, Phys.\ Lett.\ {\bf B427} (1998) 161; G.~Sterman and 
M.E.\ Tejeda-Yeomans, hep-ph/0210130.

\bibitem{muw2}
S.~Moch, P.~Uwer and S.~Weinzierl,
hep-ph/0207043 and these proceedings.


\bibitem{ancont}
T.~Gehrmann and E.~Remiddi,
Nucl.\ Phys.\  {\bf B640} (2002) 379.


\bibitem{onel}
Z.\ Bern, L.J.\ Dixon, D.C.\ Dunbar and D.A.\ Kosower,
Nucl.\ Phys.\ {\bf B425} (1994) 217;
D.A.\ Kosower, Nucl.\ Phys.\ {\bf B552} (1999) 319;
D.A.~Kosower and P.~Uwer, Nucl.\ Phys.\ {\bf B563} (1999) 477;
Z.\ Bern, V.\ Del Duca and C.R.\ Schmidt, Phys.\ Lett.\ {\bf B445}
(1998) 168;
Z.\ Bern, V.\ Del Duca, W.B.\ Kilgore and C.R.\ Schmidt, Phys.\ Rev.\
{\bf D60} (1999) 116001;
S.\ Catani and M.\ Grazzini, Nucl.\ Phys.\  {\bf B591} (2000) 435.



\bibitem{twot}
J.M.\ Campbell and E.W.N.\ Glover, 
Nucl.\ Phys.\ {\bf B527} (1998) 264;
S.\ Catani and M.\ Grazzini, Phys.\ Lett.\ {\bf B446} (1999) 143;
Nucl. Phys. {\bf B570} (2000) 287;
F.A.\ Berends and W.T.\ Giele, Nucl.\ Phys.\ {\bf B313} (1989) 595.



\bibitem{kramer}
K.~Fabricius, I.~Schmitt, G.~Kramer and G.~Schierholz, 
Z.~Phys.\ {\bf C11} (1981) 315;
W.T.~Giele and E.W.N.~Glover,
Phys.\ Rev.\  {\bf D46} (1992) 1980.



\bibitem{ert}
R.K.~Ellis, D.A.~Ross and A.E.~Terrano, Nucl.~Phys.~{\bf B178} (1981) 
421;
S.~Catani and M.H.~Seymour,
Nucl.\ Phys.\  {\bf B485} (1997) 291.


\bibitem{josnew}
S.~Moch, J.~A.~Vermaseren and A.~Vogt,
hep-ph/0209100 and these proceedings.


\bibitem{ggamma}
A.~Gehrmann-De Ridder, T.~Gehrmann and E.W.N.~Glover,
Phys.\ Lett.\  {\bf B414} (1997) 354;
A.~Gehrmann-De Ridder and E.W.N.~Glover, Nucl.~Phys.\ {\bf B517} (1998) 269.



\bibitem{4jnum}
L.J.\ Dixon and A.\ Signer,
Phys.\ Rev.\ Lett.\  {\bf 78} (1997) 811;
Phys.\ Rev.\ {\bf D56} (1997) 4031;
Z.\ Nagy and Z.\ Trocsanyi,
Phys.\ Rev.\ Lett.\  {\bf 79} (1997) 3604;
J.M.\ Campbell, M.A.\ Cullen and E.W.N.\ Glover,
Eur.\ Phys.\ J.\ {\bf C9} (1999) 245;
S.~Weinzierl and D.A.~Kosower,
Phys.\ Rev.\ {\bf D60} (1999) 054028.

\bibitem{nagy1}
Z.~Nagy and Z.~Trocsanyi,
Phys.\ Rev.\ Lett.\  {\bf 87} (2001) 082001.

\bibitem{nagy2}
Z.~Nagy,
Phys.\ Rev.\ Lett.\  {\bf 88} (2002) 122003.

\bibitem{mcfm}
J.~Campbell and R.K.~Ellis,
Phys.\ Rev.\  {\bf D65} (2002) 113007.

\end{thebibliography}
\end{document}